\documentclass[%
aip,
rsi,%
amsmath,amssymb,
reprint,%
]{revtex4-1}

\usepackage{graphicx}
\usepackage{dcolumn}
\usepackage{bm}
\usepackage{algorithm}%
\usepackage{algorithmicx}%
\usepackage{algpseudocode}%
\usepackage{float}
\usepackage[colorlinks=true, linkcolor=blue, citecolor=blue, urlcolor=blue]{hyperref}

\begin{document}
	
	\title{Experimental reservoir computing with diffractively coupled VCSELs}
	
\author{Moritz Pfl{\"u}ger}
\address{Instituto de Física Interdisciplinar y Sistemas Complejos, IFISC (UIB-CSIC), Campus Universitat de les Illes Balears, Ctra. de Valldemossa km. 7.5, 07122 Palma, Spain}
\email{moritz@ifisc.uib-csic.es}

\author{Daniel Brunner}
\address{FEMTO-ST Institute/Optics Department, CNRS - University Franche-Comté, 15B avenue des Montboucons, Besançon Cedex, 25030, France}

\author{Tobias Heuser}
\address{Institut f\"ur Festk\"orperphysik, Technische Universit\"at Berlin, Hardenbergstraße 36, 10623 Berlin, Germany}

\author{James A. Lott}
\address{Institut für Festkörperphysik, Technische Universität Berlin, Hardenbergstraße 36, 10623 Berlin, Germany}

\author{Stephan Reitzenstein}
\address{Institut für Festkörperphysik, Technische Universität Berlin, Hardenbergstraße 36, 10623 Berlin, Germany}

\author{Ingo Fischer}
\address{Instituto de Física Interdisciplinar y Sistemas Complejos, IFISC (UIB-CSIC), Campus Universitat de les Illes Balears, Ctra. de Valldemossa km. 7.5, 07122 Palma, Spain}

	\date{\today}
	
	\begin{abstract}
		
We present experiments on reservoir computing (RC) using a network of vertical-cavity surface-emitting lasers (VCSELs) that we diffractively couple via an external cavity.
Our optical reservoir computer consists of 24 physical VCSEL nodes.
We evaluate the system's memory and solve the 2-bit XOR task and the 3-bit header recognition (HR) task with bit error ratios (BERs) below 1\,\%  and the 2-bit digital-to-analog conversion (DAC) task with a root-mean-square error (RMSE) of 0.067.	
		
	\end{abstract}
	
	\maketitle
	
\textbf{Introduction.}
Research into non-classical hardware for brain-inspired computing has gained attention in recent years.
Photonic platforms have been exhibiting great potential due to the possibility of realizing high bandwidth, energy efficiency, and exploitation of the inherent parallelism of optics \cite{ChenZaijun2023,Estebanez2020,Lin2018}.

In our approach, we combine the concept of diffractive coupling (DC) \cite{ChenZaijun2023,ChenYitong2023,Lin2018,Luo2023,Bueno2017} with vertical-cavity surface-emitting lasers (VCSELs).
DC offers parallelism and the potential for highly energy-efficient implementations of neural networks (NNs).
Experimental implementations using DC include combined optical-electronic analog computing \cite{ChenYitong2023}, diffractive deep NNs \cite{Lin2018,Luo2023}, coherent VCSEL NNs \cite{ChenZaijun2023}, and reservoir computing (RC) \cite{Bueno2017}.
VCSELs are used in various experimental realizations for neuro-inspired information processing that have emerged recently \cite{Robertson2020,Zhang2021,ChenZaijun2023,Porte2021,Julian2021,Gu2022}.
VCSELs can serve to emulate a single neuron's spiking behavior \cite{Robertson2020,Zhang2021}, as nodes of optical deep NN architectures \cite{ChenZaijun2023}, or in RC implementations \cite{Porte2021,Julian2021,Gu2022}.
The conceptional simplicity of RC \cite{JaegerESN,Maass2002,Jaeger2004} allows implementing large-scale photonic NNs with current or near-term technology, and serves as an ideal springboard to investigate more complex schemes involving further optimization.
Many photonic RC implementations leverage high dimensionality based on time-multiplexing inside a long external cavity \cite{Appeltant2011}.
In these approaches, upscaling the network decreases the processing speed.

Here, we present an approach that exploits optical parallelism using a network of 24 coupled VCSELs for RC, and in which every VCSEL corresponds to one reservoir node, thus avoiding a speed penalty for time-multiplexing.
Our scheme is based on DC in an external cavity \cite{Brunner2015,VCSELcoupling,Maktoobi2019}, which has been shown to be scalable to many more emitters.
Although for individually electrically addressable VCSELs, limitations in the electrical contact design prevent a substantial increase of the here demonstrated network size, our experiments pave the way for networks with many more nodes based on micro- or nanoscale emitters, e.g. quantum-dot micropillar lasers (QDMPLs) \cite{Heuser2018}.
When testing our system on several benchmark tasks, we obtain reservoir performances that are comparable to other recent studies that use reservoirs with a similar number of physical nodes \cite{Porte2021,MaChonghuai2023}.

\begin{figure}[ht]
	\centering
	\includegraphics[width=\linewidth]{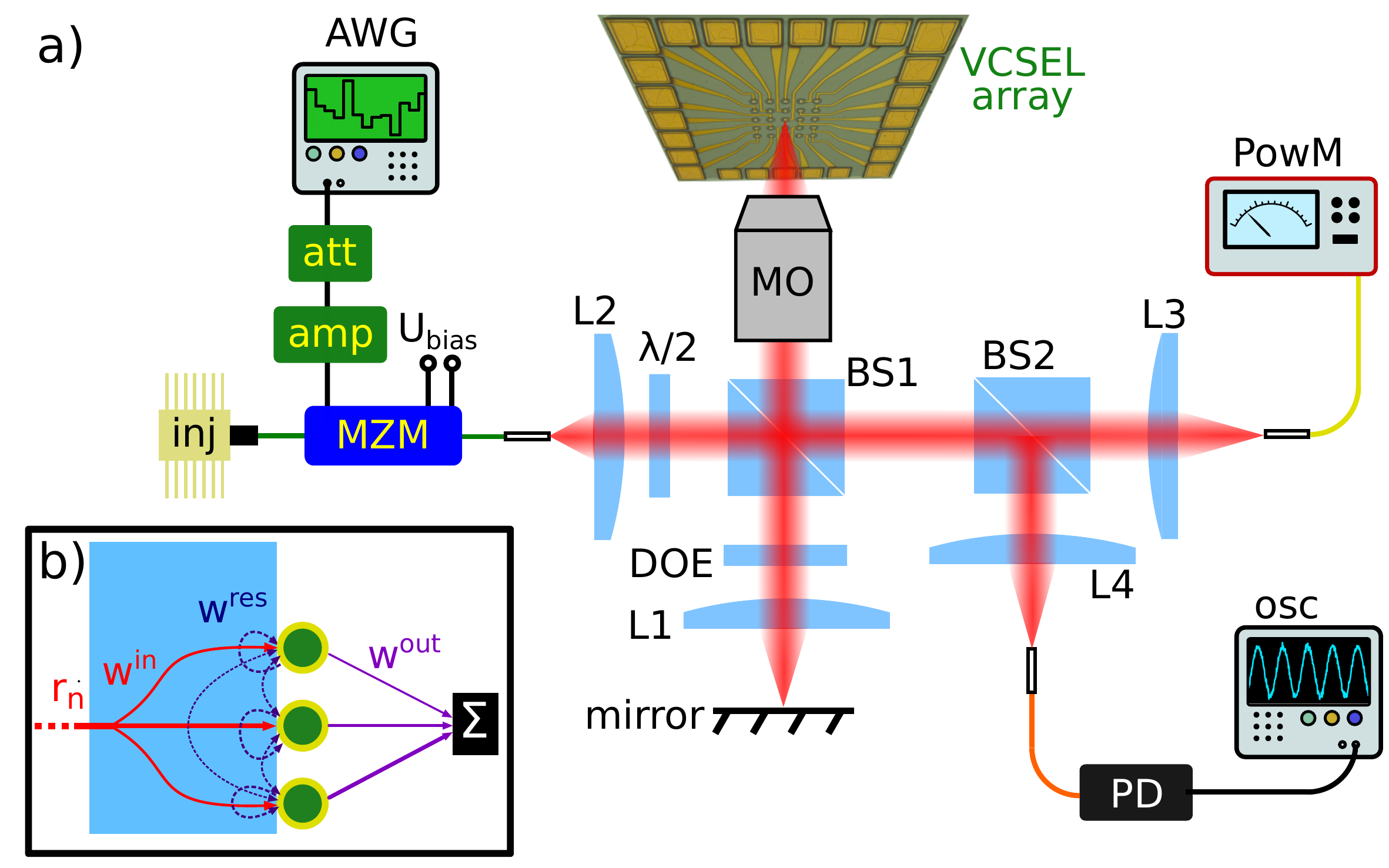}
	\caption{a) Scheme of the experimental setup. For full description, see text. Abbreviations not introduced in the text: PD = photodiode, att = attenuator, amp = radio-frequency amplifier, L1/L2/L3/L4 = lenses, PowM = powermeter for alignment, MO = microscope objective. b) Reservoir computing (RC) scheme. The light blue area indicates the external cavity, and the green circles represent the VCSELs. Pseudo-random inputs $r_n$ can be binary (bits) or continuous values between zero and one. The input weights, weights of reservoir-internal connections, and output weights are denoted by $\mathbf{w}^\mathrm{in}$, $\mathbf{w}^\mathrm{res}$, and $\mathbf{w}^\mathrm{out}$, respectively.}
	\label{fig:exp-setup-simple}
\end{figure}

\textbf{Experimental setup and reservoir computing (RC) scheme.}
In this study, we explore the possibility of using a previously demonstrated large network of diffractively coupled VCSELs \cite{VCSELcoupling} for RC.
This setup and how it can be employed as a reservoir computer is illustrated in Fig.\,\ref{fig:exp-setup-simple}.

At the heart of every reservoir computer lies the so-called reservoir, a network with fixed recurrent connections.
As nodes of our reservoir, we use custom-manufactured GaInAs quantum-well VCSELs with a slightly elliptical cross-section for polarization control, arranged in $5 \times 5$ square lattice arrays \cite{Heuser2020}.
Each VCSEL's pump current is individually controllable, and the array is homogeneous regarding emission wavelengths ($\pm 0.1\,\mathrm{nm}$ at the respective thresholds), main polarization axes ($\sigma = 4.4^\circ$), and thresholds (323\,$\mu$ A $\pm$ 26\,$\mu$ A when excluding 3 outliers) of the individual VCSELs \cite{VCSELcoupling}.
The connections between VCSELs are established via DC in an external cavity using a diffractive optical element (DOE) \cite{Brunner2015,Maktoobi2019}.
This way, every VCSEL is subject to self-feedback and bidirectional coupling to its nearest and second-nearest neighbors.
The coupling strengths between VCSEL pairs, corresponding to the weights $\mathbf{w}^\mathrm{res}$ of the reservoir-internal connections, decrease with the lattice distance between the two VCSELs.
In our experiments, we spectrally align the VCSELs with 0.01\,nm accuracy by adjusting their pump currents individually.
Previously, we demonstrated that 22 out of 25 VCSELs can be mutually optically locked based on the bidirectional coupling between the individual VCSELs \cite{VCSELcoupling}.

To inject the VCSEL array reservoir with information, we use a distributed Bragg reflector edge-emitting injection laser (inj) with an integrated optical isolator, and modulate its output intensity using an arbitrary waveform generator (AWG) and a Mach-Zehnder modulator (MZM).
The laser's output is collimated using an aspheric lens (L2), and is polarization-aligned to the VCSELs using a half-wave plate ($\lambda/2$), before it enters the external cavity via reflection at a 50/50 beam splitter (BS1).
Split up at the DOE, the light is injected into all the VCSELs simultaneously, with input weights $\mathbf{w}^\mathrm{in}$ that are strongest for the central VCSEL of the array and decrease with increasing lattice distance from the center.
When spectrally aligning the injection laser, we achieved simultaneous optical injection locking of 22 VCSELs to the injection laser \cite{VCSELcoupling}.

Part of the VCSELs' output is reflected at BS1, and is split again at a 70R/30T beam splitter (BS2), with one part being used for alignment and the other for the reservoir output layer.
The position of a multimode fiber (MMF) is adjusted to capture the signal of one VCSEL at a time, which is recorded on a 16\,GHz real-time oscilloscope (osc).
The position of the MMF was readjusted to sequentially capture the response of all 23 VCSELs that serve as output nodes.
Due to technical problems, one corner VCSEL was switched off in our experiments, and we could not record the response of the central VCSEL of the array without cross-talk from the injection laser.
Still, the central VCSEL is part of the reservoir that maps the input data onto a higher dimensional space.
We combine the individually recorded signals linearly on a conventional computer to obtain the reservoir output.
For this, we have to define the correspondence between the states of our dynamical system and the reservoir state matrix $\mathbf{Q}$, and we need to obtain the readout weights $\mathbf{w}^\mathrm{out}$ via a process called training, since the output vector $\mathbf{a}$ is given by 
\begin{equation}
	\mathbf{a} = \mathbf{Q} \mathbf{w}^\mathrm{out}.
\end{equation}
We define $\mathbf{Q} \in \mathbb{R}^{N \times J}$ via its entries $q_{n,j}$ at the $n^\mathrm{th}$ row and $j^\mathrm{th}$ column, where $q_{n,j}$ is the state of the $j^\mathrm{th}$ node of the reservoir at the $n^\mathrm{th}$ time step and $J$ and $N$ are the total number of reservoir nodes and time steps, respectively.
How we obtain discrete outputs per time step from a continuously evolving dynamical system is described in the next section.
In offline training schemes, like the one used here, $\mathbf{w}^\mathrm{out} \in \mathbb{R}^J$ is usually obtained in a single shot via ridge regression as
\begin{equation}
	\mathbf{w}^\mathrm{out} = \underset{\mathbf{v} \in \mathbb{R}^J}{\mathrm{argmin}} \left( \left\lVert \mathbf{y} - \mathbf{Q} \mathbf{v} \right\rVert^2 + \alpha \left\lVert \mathbf{v} \right\lVert^2 \right),
\end{equation}
where $\mathbf{y} \in \mathbb{R}^N$ is the target output vector, the scalar $\alpha > 0$ (here $\alpha = 0.1$) is a parameter that favors smaller weights, and $\mathbf{w}^\mathrm{out}$ is the vector $\mathbf{v}$ that minimizes the expression in parentheses.
In online experiments including real-time optimization, $\mathbf{w}^\mathrm{out}$ is usually optimized using black-box optimization concepts \cite{Bueno2017,Porte2021}, which could be implemented here by placing a spatial light modulator (SLM) in the output beam path.

\begin{figure}[ht]
	\centering
	\includegraphics{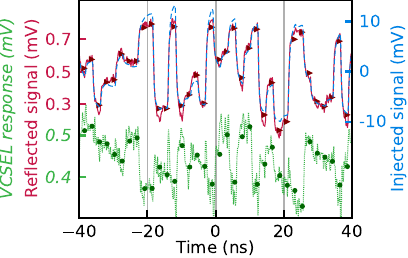}
	\caption{Injection laser time trace, directly measured (blue dashed line) and after reflection at the surface of one VCSEL (red solid line), and response of the same VCSEL (green dotted line). The green dots (red triangles) represent $q_{n,j}$ -- the state of the $j^\mathrm{th}$ node of the reservoir at the $n^\mathrm{th}$ step -- for the responses (reflections). The VCSEL implements the nonlinear transformation required for RC.}
	\label{fig:osci-resp-vcsels-a}
\end{figure}

\textbf{Dynamic response to intensity-modulated injection.}
Via modulation of the injection laser's intensity, we determined the VCSEL array's response to dynamical injection.
Sequences of 1000 uniformly distributed pseudo-random numbers $r_n$ are repeatedly injected at 454.5454\,MSamples/s, for which one value $r_n$ is injected during one external cavity roundtrip time $\tau_\mathrm{ext} = 2.2\,\mathrm{ns}$.
We account for the $\sin^2$ nonlinearity of the MZM such that the modulated intensity corresponds to $r_n$, and we adjust the amplification of the AWG output and the bias current of the MZM such that the modulation range is maximal.
We record the signals at 5\,GSamples/s and average over 1024 responses of the VCSELs to the same injected sequence to improve the signal-to-noise ratio (SNR).
In Fig.\,\ref{fig:osci-resp-vcsels-a}, the modulated intensity of the injection laser is plotted in blue and shifted by one $\tau_\mathrm{ext}$.
Its linear reflection at the surface of a deactivated VCSEL is plotted in red.
The dashed green line shows how the same VCSEL responds to the optical injection, when all the VCSELs are activated and lasing.
With the given rates for sampling and modulation, 11 sample points are recorded per $r_n$.
To obtain the reservoir state, we discard the first 2 samples due to their transient nature and define the arithmetic mean of the other 9 as $q_{n,j}$.
The $q_{n,j}$ that we obtain for the VCSEL node responses (reflections) are represented in Fig.\,\ref{fig:osci-resp-vcsels-a} by green dots (red triangles).
While the reflections follow the injected signal linearly, the VCSEL responds nonlinearly.

\begin{figure}[ht]
	\centering
	\includegraphics{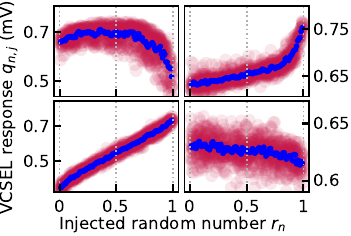}
	\caption{Dynamic responses of four different VCSELs to the injection of uniformly distributed pseudo-random numbers $r_n$ with the injection laser. Every red circle represents the response $q_{n,j}$ to the injection of $r_n$, the blue dots are averages.}
	\label{fig:osci-resp-vcsels-b}
\end{figure}

In Fig.\,\ref{fig:osci-resp-vcsels-b}, the $q_{n,j}$ are plotted against the injected pseudo-random numbers $r_n$, and the blue data are obtained from a  0.02-wide sliding window average.
We observe a large variety of responses for different VCSELs.
Importantly, some of the VCSELs respond nonlinearly (see upper two panels), thus fulfilling this prerequisite for RC.



\textbf{Basic benchmark tasks.}
We tested our VCSEL array RC on four basic benchmark tasks: 1) memory capacity (MC); 2) header recognition (HR); 3) the exclusive or (XOR); and 4) digital-to-analog conversion (DAC).
For MC, the input $r_n$ consists of uniformly distributed pseudo-random numbers, while for HR, XOR, and DAC, pseudo-random bits are injected.
For training and testing, we use 5-fold cross validation \cite{Arlot2010}.
In our experiments, we scanned two main parameters to find the best operating point.
First, we can use neutral density filters to reduce the average power ratio
\begin{equation}
	\epsilon = \frac{1}{J} \sum_{j=1}^J \frac{\left.P_\mathrm{inj} \right|_{\mathrm{tf}(j)}}{P_j},
\end{equation}
where $\left. P_\mathrm{inj} \right|_{\mathrm{tf}(j)}$ is the power of the injection laser at the top facet of VCSEL $j$ and $P_j$ is the optical intensity emitted by VCSEL $j$.
Second, we can control the wavelength detuning $\Delta \lambda$ between the injection laser and the average VCSEL wavelength via the injection laser temperature $T_\mathrm{inj}$.

\begin{figure}[htb]
	\centering 
	\includegraphics{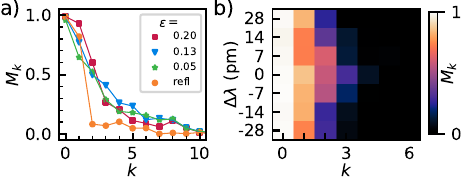}
	\caption{Memory correlation $M_k$ vs. number of time steps back $k$ (a) for $\Delta \lambda = 0{pm}$ with different power ratios $\epsilon$ and for the linear reflections (refl.) at the VCSELs' surfaces and (b)  for $\epsilon = 0.20$ with different detuning $\Delta \lambda$ of the injection laser from the average VCSEL emission wavelength.}
	\label{fig:mem-corr}
\end{figure}

To quantify our reservoir's memory, we determine the memory correlations $M_k$.
For this, the target output is the value injected $k$ steps before, i.e. $y_n = r_{n-k}$, and the output weights are determined separately for every $k$.
We calculate $M_k$ as defined by Jaeger \cite{JaegerMC}
\begin{equation}
	M_k = \left( \frac{\sum_n (a_{n,k}-\overline{a})(r_{n-k}-\overline{r})}{\sigma_a \sigma_r} \right)^2,
\end{equation}
where $\overline{a}$ and $\overline{r}$ are the means of the outputs $a_n$ and inputs $r_n$, respectively, and $\sigma_a$ and $\sigma_r$ are their standard deviations.
In Fig.\,\ref{fig:mem-corr}, $M_k$ is shown for different $\epsilon$ and $\Delta \lambda$.
We observe that slightly attenuating the injection increases $M_k$ for $k \geq 3$, while detuning the injection laser from the VCSELs reduces it.
Surprisingly, using the states $q_{n,j}$ obtained from the reflections gives $M_1>0.8$.
The reason for this is probably that the reflection at the VCSEL surfaces contributes to the recorded signal after one additional round-trip through the external cavity.
Summing up the $M_k$ for $k \leq 10$, we obtain memory capacities of up to 3.6.
We do not include $M_k$ for $k > 10$ in the sum to avoid summing principally negligible distributions that could be generated by noise.

\begin{figure}[htb]
	\centering
	\includegraphics{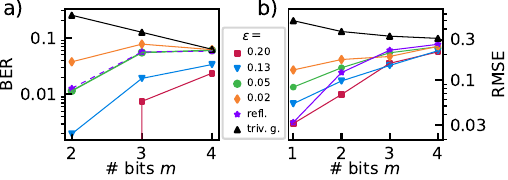}
	\caption{a) Bit error ratios (BERs) for the 2-bit, 3-bit and 4-bit header recognition (HR) task. b) Root-mean-square error (RMSE) for 1- to 4-bit digital-to-analog conversion (DAC). Both for different power ratios $\epsilon$, for the reflections (refl.), and for the trivial guess (triv. g.). Lines are guides to the eye.}
	\label{fig:HRyDAC}
\end{figure}

For the $m$-bit HR task, we first pick a target sequence consisting of $m$ bits.
If the bits from $r_{n-m+1}$ to $r_n$ correspond to this target sequence, $y_n = 1$, otherwise $y_n = 0$.
Since the targets for this task are Boolean, the reservoir output $a_n$ is thresholded.
Averaging the bit error ratios (BERs) over all $2^m$ possible $m$-bit sequences gives the overall BER for $m$-bit HR.
The dependence of the BERs on $m$ is plotted for different $\epsilon$ and for the reflections in Fig.\,\ref{fig:HRyDAC}\,a).
Trivially guessing zeros yields BER$_0=2^{-m}$, marked by the black triangles.
Note that the task requires memory, since the classification occurs based only on the current state of the reservoir ($q_{n,j}$).
In general, the BER decreases with increasing $\epsilon$.
This might be due to the low SNR and low MC for low $\epsilon$.
Using the reflections, and thus a mostly linear reservoir, yields comparable results to $\epsilon = 0.05$.

XOR tasks are popular nonlinear benchmark tasks. For the $m$-bit XOR task, the target output is
\begin{equation}
	y_n = \left( \sum_{n-m+1}^n r_n \right) \mathrm{mod}\,2,
\end{equation}
and, as for HR, the reservoir output is thresholded.
The BERs for XOR are shown in Tab.\,\ref{tab:BER-XOR}.
Again, we observe that the BER increases with decreasing $\epsilon$.
The results also reflect the substantial increase in difficulty from the 2-bit to the 3-bit XOR task.
Similar to HR, using the reflections yields a good performance.

\begin{table}[htb]
	\centering
	\caption{\bf BERs for the 2-bit and 3-bit XOR task at different power ratios $\epsilon$ and for using the reflections.}
	\begin{tabular}{c|ccccc}
		\hline
		$\epsilon$ & 0.20 & 0.13 & 0.05 & 0.02 & refl. \\
		\hline
		BER 2-bit XOR & 0.008 & 0.023 & 0.158 & 0.258 & 0.065 \\
		BER 3-bit XOR & 0.257 & 0.318 & 0.400 & 0.462 & 0.355 \\
		\hline
	\end{tabular}
	\label{tab:BER-XOR}
\end{table}

For $m$-bit DAC, the aim is to map the $2^m$ different bit sequences consisting of the bits from $r_{n-m+1}$ to $r_n$ onto their normalized analog value in the interval $[0;1]$.
Considering the first injected bit as the most significant bit leads to
\begin{equation}
	y_n = \sum_{k=1}^{m} \frac{2^{m-k} r_{n-m+k}}{2^m-1}.
\end{equation}
We measure the performance by calculating the root-mean-square error
\begin{equation}
	\mathrm{RMSE} = \sqrt{\frac{1}{N} \sum_{n=1}^N (a_n-y_n)^2}.
\end{equation}
The results for different $\epsilon$ and for the reflections are plotted semilogarithmicly against $m$ in Fig.\,\ref{fig:HRyDAC}\,b).
For comparison, we give the expected RMSE for trivially guessing $a_n = 0.5$ for all $n$.

\textbf{Discussion.}
We tested our optical RC on basic benchmark tasks and obtained a maximal MC of 3.6, a minimal BER of 0.008 for the 2-bit XOR task, error-free 2-bit HR, a minimal BER of 0.007 for 3-bit HR, and 2-bit DAC with a minimal RMSE of 0.067.
This is comparable to results obtained in RC with similar numbers of nodes \cite{Porte2021,MaChonghuai2023}.

An SLM could be used to record the weighted output of all the VCSELs simultaneously \cite{Porte2021}.
By implementing this optical output layer, one would additionally increase the SNR due to adding signals with uncorrelated noise sources.
Replacing the cavity mirror by an SLM would, in addition, allow reconfiguring the connections between nodes.
Since we obtain better results for high injection power ratios $\epsilon$ for all tasks, we assume that the experiment could benefit from an injection laser with higher output power.
This computing scheme could also be combined with the delay-based approach to increase the number of reservoir nodes \cite{Sugano2020}.
Finally, the coupling scheme can be extended to many more nodes \cite{Maktoobi2019}.
For this, arrays of QDMPLs \cite{Heuser2018} could replace the VCSEL arrays, which brings about new challenges, like operation at cryogenic temperatures and optical pumping.


\textbf{Conclusion.}
We employ diffractively coupled VCSELs in an array for processing information using the concept of RC.
Our reservoir consists of 24 physical nodes that are represented by the VCSELs, of which 23 contribute to the reservoir output.
The reservoir-internal connections are established via diffraction in an external cavity.
This diffraction scheme additionally allows for injecting information into all the nodes simultaneously.
The output layer is implemented by recording the VCSELs' responses individually and linearly combining them on a conventional computer.
By analyzing the VCSELs' dynamic response to intensity-modulated injection, we confirm that nonlinear transformations take place, which is a prerequisite for RC.
We find that our RC has high 1-step memory, but this quickly fades to $M_k < 0.5$ for $k \geq 3$ and to $M_k < 0.2$ for $k \geq 6$.
Correspondingly, we solve the $m$-bit versions of four basic benchmark tasks with low errors or error-free for $m = 2$, but upon increasing $m$, the errors increase rapidly.
Our results serve as a proof of concept for RC with diffractively coupled lasers.
We expect that more complex tasks could be solved with larger networks.
These could be established either by combining our approach with a delay-based scheme or by using other types of lasers.


\textbf{Funding}
Agencia Estatal de Investigación (CEX2021-001164-M); Volkswagen Foundation (NeuroQNet I, NeuroQNet II); Deutsche Forschungsgemeinschaft (SFB 787).
	
\textbf{Acknowledgments}
M.P. would like to thank Apostolos Argyris and Irene Estébanez Santos for help in the laboratory and fruitful discussions.
 This work was partially supported by the Program for Units of Excellence in R\&D María de Maeztu (CEX2021-001164-M	/10.13039/501100011033).

\section*{References}
\bibliography{mybib}
	
\end{document}